\documentclass[prb,reprint,superscriptaddress,amsmath,amssymb,floatfix]{revtex4-1}

\usepackage{graphicx}
\usepackage{dcolumn}
\usepackage{multirow}
\usepackage{bm}
\usepackage{wrapfig}

\begin{document}

\title{Magnetic properties of YCo$_5$ compound at high pressure}
\author{E. Burzo}
\affiliation{Faculty of Physics, Babes-Bolyai University 40084 Cluj-Napoca, Romania}
\affiliation{Romanian Academy of Science, Cluj-Napoca Branch, Cluj-Napoca 400015, Romania}
\author{P. Vlaic}
\affiliation{University of Medicine and Pharmacy “Iuliu Hatieganu”, Physics and Biophysics Department Cluj-Napoca, Romania}
\author{D. P. Kozlenko}
\affiliation{Frank Laboratory of Neutron Physics, Joint Institute for Nuclear Research, 141980 Dubna, Russia}
\author{N.~O.~Golosova}
\affiliation{Frank Laboratory of Neutron Physics, Joint Institute for Nuclear Research, 141980 Dubna, Russia}
\author{S.~E.~Kichanov}
\affiliation{Frank Laboratory of Neutron Physics, Joint Institute for Nuclear Research, 141980 Dubna, Russia}
\author{B.~N.~Savenko}
\affiliation{Frank Laboratory of Neutron Physics, Joint Institute for Nuclear Research, 141980 Dubna, Russia}
\author{A. \"Ostlin}
\affiliation{Theoretical Physics III, Center for Electronic Correlations and Magnetism, Institute of Physics, University of Augsburg, D-86135Augsburg, Germany}
\author{L. Chioncel}
\affiliation{Theoretical Physics III, Center for Electronic Correlations and Magnetism, Institute of Physics, University of Augsburg, D-86135Augsburg, Germany}
\affiliation{Augsburg Center for Innovative Technologies (ACIT),D-86135 Augsburg, Germany}

\date{\today}

\begin{abstract}
The crystal structure and magnetic properties of YCo$_5$ compound have 
been studied by neutron diffraction, in the pressure range 
$0 \le p \le 7.2 \ GPa$. The experimental data are analyzed together with
results from the combined Density Functional and Dynamical Mean-Field 
Theory. A good agreement between the experimentally determined and 
calculated values of cobalt moments is shown. Our scenario 
for the behavior of YCo$_5$ under pressure, is the combined action 
of the Lifshitz transition with a strong local electron-electron 
interaction.
\end{abstract}

\pacs{Valid PACS appear here}
\maketitle


\section{\label{sec:intro}Introduction}

The YCo$_5$ compound crystallize in the CaCu$_5$-type structure, 
space group $P6/mmm$, with the cobalt atoms occupying $2c$ and $3g$ 
sites, while yttrium is located at the $1a$ site. 
The unit cell is formed of alternating YCo($2c$)/Co($3g$) layers.
The compound is 
ferromagnetic with cobalt moments of a dominant $3d$-character. 
The magnetic moment at the Co($2c$) site is a little higher than that 
at the Co($3g$) site. Yttrium has an induced magnetic moment of 
$4d$-band character which is significantly smaller and  
negatively polarized~\cite{bu.ch.90}.
First-principle, Density Functional 
Theory calculations were performed  to investigate the pressure 
effects upon the crystal structure and magnetic properties of YCo$_5$~\cite{ro.ko.06,ko.sc.08,bu.vl.13}.
The hallmark of the electronic structure in the YCo$_5$ 
compound 
is the presence of majority spin Co flat-bands of $3d$-character, 
resulting from dispersion-less (localized) eigenstates along particular
directions~\cite{ma.ar.15} in the Brillouin zone. 
Upon pressure, electronic structure calculations predict a topological 
change in the band structure: the majority spin Co-$3d$ flat-band, 
located 
below $E_F$ at ambient pressure, is shifted upwards in energy 
and leads to a structural instability when promoted above the Fermi 
level~\cite{ro.ko.06,ko.sc.08}. This isomorphic structural transformation
of YCo$_5$ was ascribed~\cite{ro.ko.06,ko.sc.08} to a first order 
Lifshitz transition~\cite{lifs.60}. The corresponding threshold pressure
for the electronically topological transition (ETT) to happen is predicted 
to be in the range of $10-20$~GPa (Fig.~2, Ref.~\onlinecite{ro.ko.06}),
depending on theoretical or experimental analysis~\cite{ro.ko.06,ko.sc.08}. 
A transition from strong to weak ferromagnetism~\cite{ro.ko.06}
was also associated with the ETT.

In this paper we report results of neutron diffraction (ND) 
measurements on YCo$_5$ compound, in the pressure range $0 \le p \le 7.2~GPa$ 
and ambient temperature. To our best knowledge ND studies at higher
pressures were not performed on YCo$_5$ nor are available in the literature.
We supplement our experimental study, with results
of electronic structure calculations using the combination of Density 
Functional~\cite{kohn.99} and Dynamical Mean Field 
Theory~\cite{me.vo.89,ko.vo.04}, the LDA+DMFT~\cite{ko.sa.06,held.07} 
method. At the standard DFT(LSDA) level we confirm the pressure dependent 
band structure calculations reported previously~\cite{ro.ko.06,ko.sc.08}. 
Recent LDA+DMFT calculations for YCo$_5$~\cite{zh.ja.14} discussed the electronic
correlations effects without pressure, in particular the formation and
enhancement of orbital moments and magnetocrystalline anisotropy, as a function of
electron-electron interaction parameters. 
In the current paper we follow the same methodology and complement this previous study,
by discussing results for the band structures, density of states.  
The scenario we propose for the pressure dependence of the physical
properties of YCo$_5$ takes into account the interplay 
between electronic correlations and Lifshitz transitions.

The paper is organized as follows: In Sec.~\ref{sec:exp_tech}
after a brief description of the experimental techniques the pressure 
dependence of neutron diffraction spectra,
the changes in the unit cell parameters and
magnetic moments is presented and analyzed. 
In Sec.~\ref{sec:dft} the experimental data is
compared with the different models of the band structures obtained within 
DFT and its LDA+DMFT extension. Sec.~\ref{sec:disc+concl} presents a
discussion and concludes our paper.

\section{Experimental techniques} 
\label{sec:exp_tech}
The YCo$_5$ compound has been prepared by melting the high purity elements in an
induction furnace, under high purity argon atmosphere. An excess of $2~\%$ yttrium 
was used in order to compensate their loss during melting. The sample has been 
thermally treated at $1050~^{\circ}C$ for 5 days. The X-ray diffraction pattern evidenced the presence of only one phase. The thermal variation of magnetization has been determined in the temperature range $4.2$- $1000 K$ and field up to $70~kOe$. 

Neutron diffraction measurements were performed in the pressure range 
$0 \le p \le 7.2~GPa$, at $T = 290 K$, with $DN$-$12$ spectrometer~\cite{ak.ba.99}
at the $IBR$-$2$ high flux pulsed reactor (FLNP JINR Dubna, Russia), using
sapphire anvil high pressure cells~\cite{gl.go.91}. 
The sample volume was about $4~mm^3$. Several tiny ruby chips were placed, at different
points, on the sample surface. The pressure was determined by ruby fluorescence
technique with the accuracy of $0.05~GPa$, at each ruby chip and the pressure on 
the sample was obtained by averaging the values determined at different points. 
Diffraction patterns were collected at scattering angles $45.5^\circ$ and $90^\circ$. 
The spectrometer resolution, at $\lambda = 2$ {\AA }, is $\Delta d/d = 0.022$ and $0.015$
for these angles, respectively. The typical data for collection time, at one temperature, was $20$ hours.

\subsection{Crystal and magnetic structures}

\begin{figure}[h]
\begin{center}
\includegraphics[width=\linewidth, clip=true]{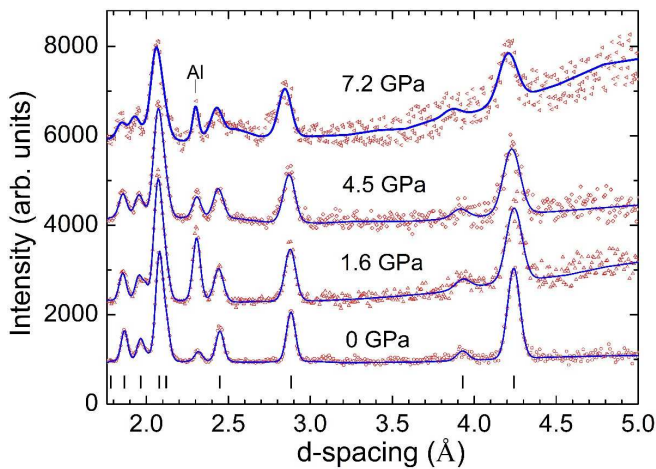} 
\end{center}
\caption{Neutron diffraction patterns of YCo$_5$ at selected 
pressures, processed by the Rietveld method. The experimental points 
(red dots) and the calculated profiles (blue solid lines) are shown. 
The diffraction peaks include both nuclear and magnetic 
contributions and the ticks below the calculated profile indicate 
the position of the maxima. The position of extra peak from Al gasket 
of high pressure cell is also shown.} 
\label{fig1}
\end{figure}

The neutron patterns were analyzed by Rietveld method, using 
MRIA~\cite{zl.ch.91} and Full prof~\cite{rodr.93} programs. 
The errors in determining the cobalt moments, particularly in the high
pressure range, are of  $0.10 \mu_B$ - Table~\ref{tab1}. These are 
higher than the differences between magnetic moments at $2c$ and $3g$ 
sites~\cite{bu.ch.90}. Consequently, even in the analysis of the high 
pressure data, the ordered magnetic moments of Co atoms at $2c$ and $3g$ 
sites, were assumed to be equal.
The neutron diffraction patterns on YCo$_5$ compound, at $T = 290 K$ and 
$p \le 7.2~GPa$, evidenced the presence of only one phase, having 
CaCu$_5$-type structure - Fig.~\ref{fig1}.

\begin{figure}[h]
\begin{center}
\includegraphics[width=0.99\linewidth, clip=true]{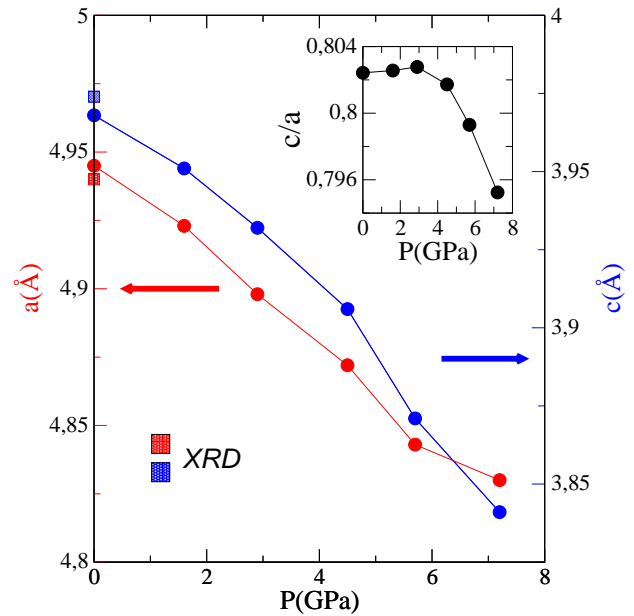}
\end{center}
\caption{The pressure dependences of lattice parameters, $a$ (red circles) and $c$ (blue circles). The variation of 
the $c/a$ ratio as a function of pressure is shown in the inset. 
The blue/red boxes represent the $c$ and $a$ values obtained by XRD 
at room temperature and ambient pressure.}
\label{fig1b}
\end{figure}

The $a$ and $c$ lattice parameters decrease with pressure, in an
anisotropic way, see  Fig.~\ref{fig1b}. The actual values
of the unit cell parameters are provided in Tab.~\ref{tab1}. The $c/a$ 
ratio (the inset of Fig.~\ref{fig1b}) increases slightly with
pressure, reaches a maximum at about $p = 2.9~GPa$ ($V/V_0 = 0.972$), 
then decreases further in value.
At the maximum available pressure used in the 
present study ($p = 7.2~GPa$) a volume reduction of   
$V/V_0 \approx 0.924$ is obtained. Note that the lattice parameters
determined by XRD analysis at ambient conditions, are in excellent 
agreement with those obtained by neutron diffraction.
At low temperatures the behavior of the  $c/a$ ratio, in the vicinity 
of the critical pressure allows to identify the nature of the transition
as of first order~\cite{ro.ko.06}. 
At $T=100$~K, there is a sharp decrease of
the $c/a$ ratio from $75.5$~{\AA}$^3$ up to a minimum value at 
$74$~{\AA}$^3$~\cite{ko.sc.08}. At $T=295$~K, the decrease of this 
ratio is nearly linear and in an extended volume range,
from $77.5$~{\AA}$^3$ to $74$~{\AA}$^3$ (Fig.~7, Ref.~\onlinecite{ko.sc.08}).
This suggest a possible change of transition type.

\begin{table*}
\begin{tabular}{c|cccccc}
\hline
$p$(GPa) & 0 & 1.6 & 2.9 & 4.5 & 5.7 & 7.2  \\
\hline \hline
$a$ ({\AA }) & 4.945(3) & 4.923(7) & 4.898(9) & 4.872(9) & 4.843(9) & 4.830(9) \\
$c$ ({\AA }) & 3.968(4) & 3.951(8) & 3.932(8) & 3.906(8) & 3.871(9) & 3.841(9) \\
\multicolumn{1}{c|}{\multirow{1}{*}{\shortstack{Mean Co \\ moment}}} & 1.48(9) &  1.36(10) & 1.35(10) & 1.29(10) & 1.24(10) &  1.05(10) \\
\hline 
\end{tabular}
\caption{\label{tab1} Measured pressure-dependent lattice parameters and mean cobalt moments.}
\end{table*}

The XRD studies under pressure in combination with density functional
electronic structure calculations on YCo$_5$ compound suggested 
that an isomorphic structure transition takes place at $p = 19~GPa$ at 
low temperature, while at room temperatures a similar transition happens 
for a pressure of $12~GPa$~\cite{ro.ko.06,ko.sc.08}. Such a transition 
was described as a first order Lifshitz transition~\cite{lifs.60}.

\begin{figure}[h]
\begin{center}
\includegraphics[width=0.95\linewidth, clip=true]{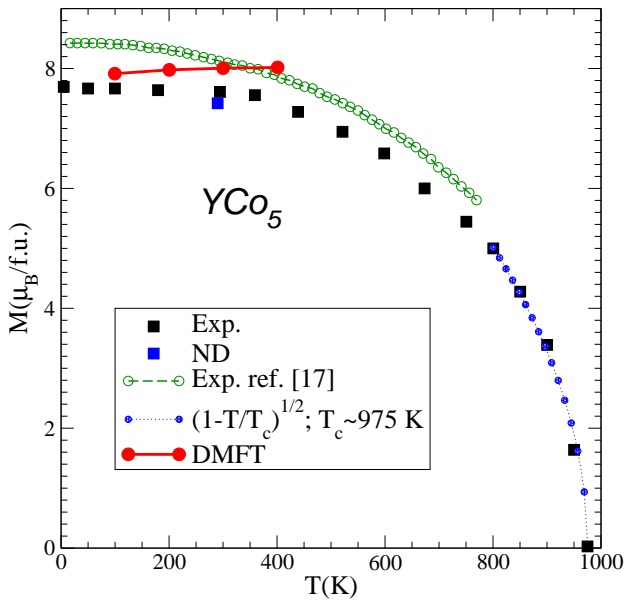}
\end{center}
\caption{Temperature dependence of spontaneous magnetization at ambient pressure.
Computed data: Red solid line - DMFT. The last few points
of the neutron diffraction data (black squares) were fitted with $\sqrt{1-T/T_c}$
(blue circles). Single-crystal experiments of Ref.~\onlinecite{ch.sa.17}, green 
circles. Difference between single-crystal data (green circles) and 
powder sample (black squares) are consistent with previous findings~\cite{da.ke.96}.}
\label{fig2}
\end{figure}

The YCo$_5$ compound orders ferromagnetically~\cite{bu.ch.90}. 
The thermal variation of magnetization at ambient pressure is given 
in Fig.~\ref{fig2}. 
At $T = 4.24$ K the saturation magnetization is $7.7~\mu_B/f.u.$. 
In the temperature range $4.2$-$300$ K, the magnetization decrease 
by $3\%$ being of $7.42~\mu_B/f.u.$, at ambient condition,
while the mean cobalt moment of $1.48~\mu_B/$atom remains nearly the same.
In order to compare the evolution with pressure of total moments, the experimentally determined values, at $T = 290$ K, were extrapolated to $T = 0$ K, according to $T^{3/2}$ law, which describes also the temperature dependence of YCo$_5$ magnetization, at $T/T_c \le 0.3$ - Fig.~\ref{fig2}. The extrapolated cobalt moments, are only by  $0.04~\mu_B$ higher than those determined at $T = 290$K. 
Finally, the measured and computed (DMFT) total moments agree rather well, see Fig.~\ref{fig2}. 
In addition we compare in Fig.~\ref{fig2} our results also with the recent experiments,
of Ref.~\onlinecite{ch.sa.17}, on single crystals.
The DMFT results show a rather uniform temperature dependence in the 
range of 200 to 600K. The fitting of the experimental data with a  $\sqrt{1-T/T_c}$ dependence in the temperature range of 800 to 1000K, leads to a value $T_c \approx 975$~K, which is
in the range of experimental values discussed in the review paper
Ref.~\onlinecite{busc.77}.

The relatively large orbital moments~\cite{de.gi.76} of Co
couple to the spin moments (spin-orbit coupling) and establish
the direction of cobalt spin moments along the crystallographic axes.
The local anisotropy (stabilization energy) was experimentally 
estimated at $2.88\cdot~10^{-4}$~erg/atom for Co($2c$) sites and a smaller
but opposite contribution $-0.84\cdot 10^{-4}$~erg/atom arises from the
Co($3g$) sites~\cite{stre.78,stre.79}. 
The anisotropy energy, can also be computed using the density functional
theory~\cite{da.ke.91,no.br.92,ya.as.96,jans.99,ya.te.91,st.ri.01}. 
Its magnitude was found to be strongly affected by changes of the 
lattice geometry ($c/a$ ratio and volume)~\cite{st.ri.01}, 
and by the degree of filling of Co-$3d$ bands. In particular for the band
filling corresponding to a mean cobalt moment of  $0.6~\mu_B$, an easy
plane of magnetization was shown to be favored~\cite{st.ri.01}.

The cobalt moments, $M_{Co}$, as function of pressure $p$, 
are given in Tab.~\ref{tab1}.  
As can be seen, the cobalt moments change little with pressure. For a decrease 
in relative volume up to $V/V_0~\approx~0.92$, the changes were
interpreted as a consequence of a high-spin to low-spin state
transition~\cite{ro.ko.06,ko.sc.08} with a 
larger decrease for the cobalt moments at the $3g$ position. This behavior was also connected
with the different local environments of Co($2c$) and Co($3g$) sites. 
By further increasing pressure, up to a relative volume $V/V_0 = 0.8$ the cobalt
moments collapse as previously obtained by full potential density functional calculations~\cite{ro.ko.06,ko.sc.08}.
Along with the high-spin to low-spin transition, the spin reorientation transition
(frequently present in the 
family or RCo$_5$ compounds~\cite{da.ke.91,no.br.92,ya.as.96,jans.99,ya.te.91,st.ri.01})
may also take place and contribute to the moment reduction.
In addition we observe that the combined structural and magnetic transformation 
resembles the situation already pointed out for the GdCo$_5$ compound~\cite{bu.vl.13}.
In the ferromagnetic GdCo$_5$ the magnetic interaction paths follow 
the Campbell model~\cite{camp.72} and because the Gd($4f$)-states are situated 
away from the Fermi level, at the Fermi surface the bands 
have a dominant Co~($3d$)-states character. Having the same crystal structure 
dispersion-less Co~($3d$)-bands are expected to be seen in the band structure~\cite{oc.ry.15} of GdCo$_5$.  
Therefore, we {\it predict} that under pressure GdCo$_5$ may also show 
signatures of a possible Lifshitz transition, unnoticed previously because
of the focus on the strong magnetism of $4f$ states.

\section{Density Functional Theory Calculations}
\label{sec:dft}

\begin{figure*}
\begin{center}
\includegraphics[scale=0.45, clip=true]{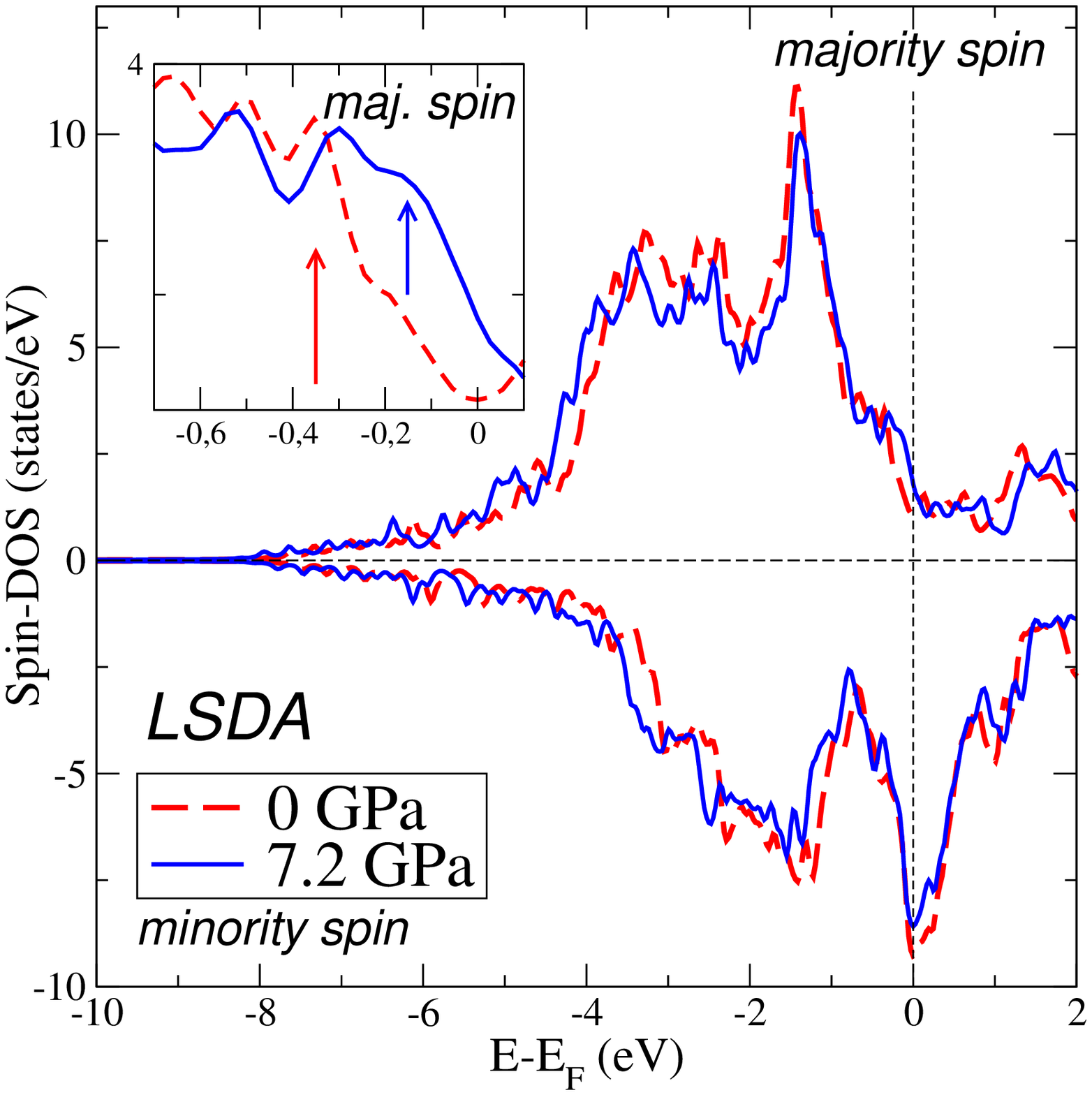}
\includegraphics[scale=0.45, clip=true]{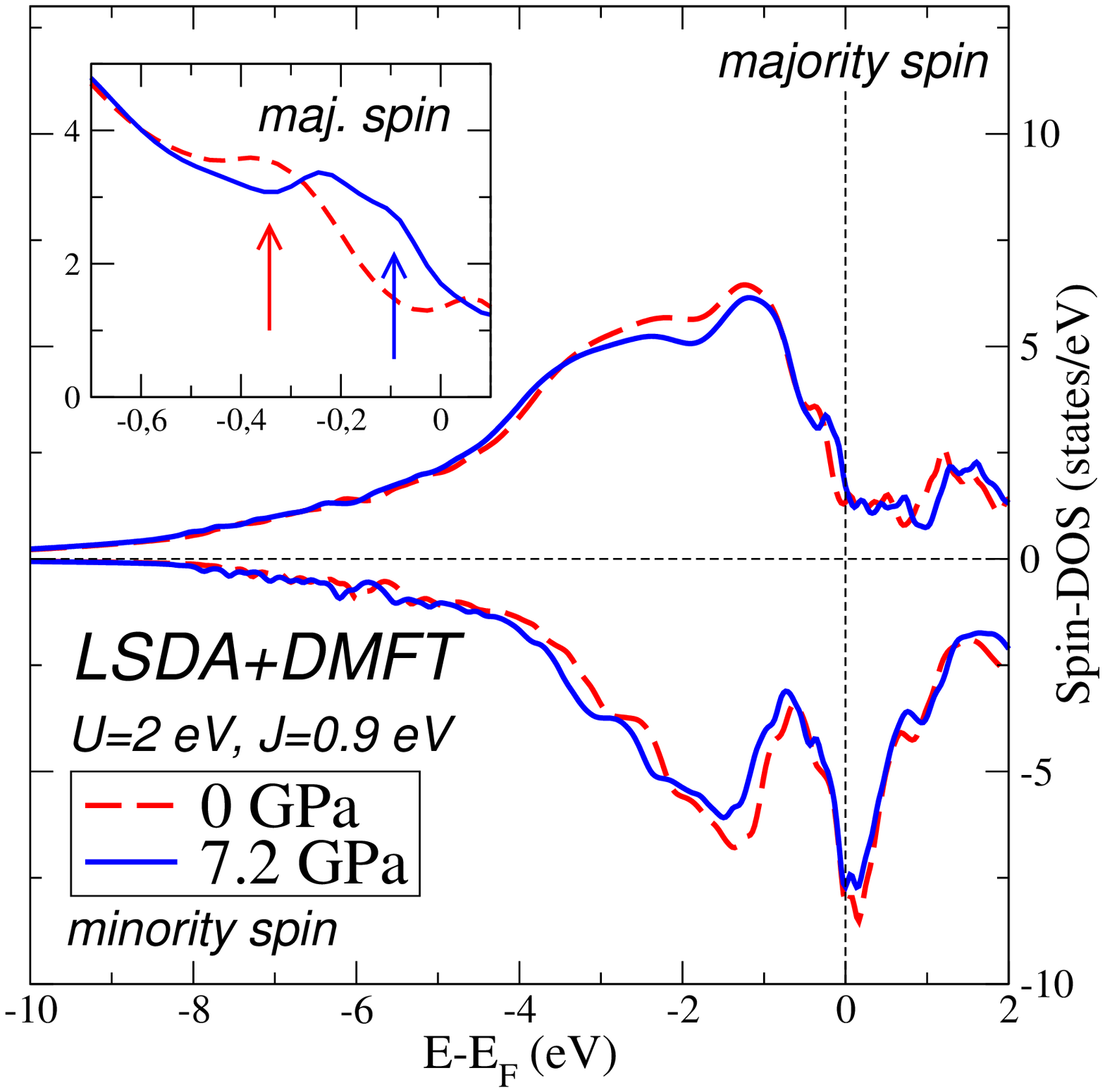}
\end{center}
\caption{Spin-resolved total densities of states for LSDA (left) and LSDA+DMFT (right).
Ambient pressure and $p=7.2$ GPa is denoted by red dashed lines and blue 
solid lines, respectively. Arrows in the insets indicate the change in the 
position of flat bands.}
\label{fig3}
\end{figure*}

In the present paper the full-potential linearized muffin-tin orbitals 
(FPLMTO) method, as implemented in the {\sc RSPt} code~\cite{rsptbook,gr.ma.12}, was
employed.
The calculations have been performed using the LDA with the parametrization of
Perdew and Wang~\cite{pe.wa.92} for the exchange-correlation functional.
Three kinetic energy tails were used, with corresponding energies $0.3$, $-2.3$,
and $-1.5$ Ry. The ${\mathbf{k}}$-mesh grid was $6 \times 6 \times 6$, and
a Fermi-Dirac smearing with $T = 474$ K was used. 
The  muffin-tin radius was set to $2.1$ a.u. for the Co atoms, and to $2.3$ a.u.
for the Y atom, and was kept constant for all pressures. The calculations 
included spin-orbit coupling and scalar-relativistic terms.
A very important property of YCo$_5$ is the magnetocrystalline anisotropy
which derives from the large orbital magnetic moments on Co observed
in experiment~\cite{stre.79,sc.ta.80,al.gi.81}. 
In order to capture the large orbital polarization (OP), 
non-ab-initio DFT+OP has been previously employed~\cite{no.er.90,be.ab.15}. 
The physical 
origin of the proposed OP corrections derives from the atomic limit. 
Alternatives to this method are the relativistic implementation of the 
LDA+U or the more recently developed LDA+DMFT. In both these techniques 
a multi-orbital Hubbard
term is added to the DFT, and the ``upgraded'' Hamiltonian is solved 
either by a static mean field decoupling (+U), or by using a full 
many-body approach 
(DMFT). The Hubbard terms are constructed from a set of Coulomb $U$ and exchange $J$ parameters.
We have performed in this paper 
LDA+DMFT calculations within the FPLMTO method.
For the DMFT scheme, the perturbative spin polarized $T$-matrix fluctuation-exchange (SPT-FLEX) 
impurity solver has been used~\cite{li.ka.97}. 
The solver is implemented on the imaginary-axis
Matsubara domain, and the temperature was set equal to that used for the Fermi-Dirac smearing.
We consider Hubbard corrections only to the Co-$3d$ orbitals, 
set $J=0.9$ eV, while allowing $U$ to vary in the range of 
$1.0$~eV to $3.0$~eV.

\subsection{Density of States and Band Structures}
In all calculations we include the SO coupling 
and take the magnetization oriented along the $c$-axis.
Nevertheless, we have checked that for the cases in which the moment is oriented 
within the ($ab$)-plane the $3g$ sites splits into two inequivalent 
sites, one of multiplicity one and the other
with multiplicity two.

In Fig.~\ref{fig3} we show the total spin-resolved density of states
using the LSDA (left panel) and including the DMFT correction
(right panel).
For both LSDA(+DMFT) density of states the majority spin channel ($\uparrow$) is 
almost complete, while the Fermi level in the minority spin ($\downarrow$)
is pinned around a maximum. 
The orbital contribution to the maximum in the minority spin DOS 
has a predominant $d_{x^2-y^2}/d_{xy}$-character. Both Co($2c$)/($3g$)-sites
contribute, however Co($2c$)-sites have a stronger weight around $E_F$. 
Correlation induced modification in DOS (seen in Fig.~\ref{fig3}) are: (i) 
a broadening of the spectra because of the presence of many-body 
self-energy $\Sigma_{}(E)$, and (ii) the appearance of
tails in the density of states at higher binding energies. These 
changes are similar for the ambient pressure ($V/V_0 = 1.0$) as 
well as for the $p=7.2$~GPa ($V/V_0 = 0.924$). 
The insets of Fig.\ref{fig3} present the majority spin-channels ($\uparrow$), 
with the arrows pointing to the energies at which the dispersion-less bands are obtained.

\begin{figure*}
\begin{center}
\includegraphics[scale=0.45, clip=true]{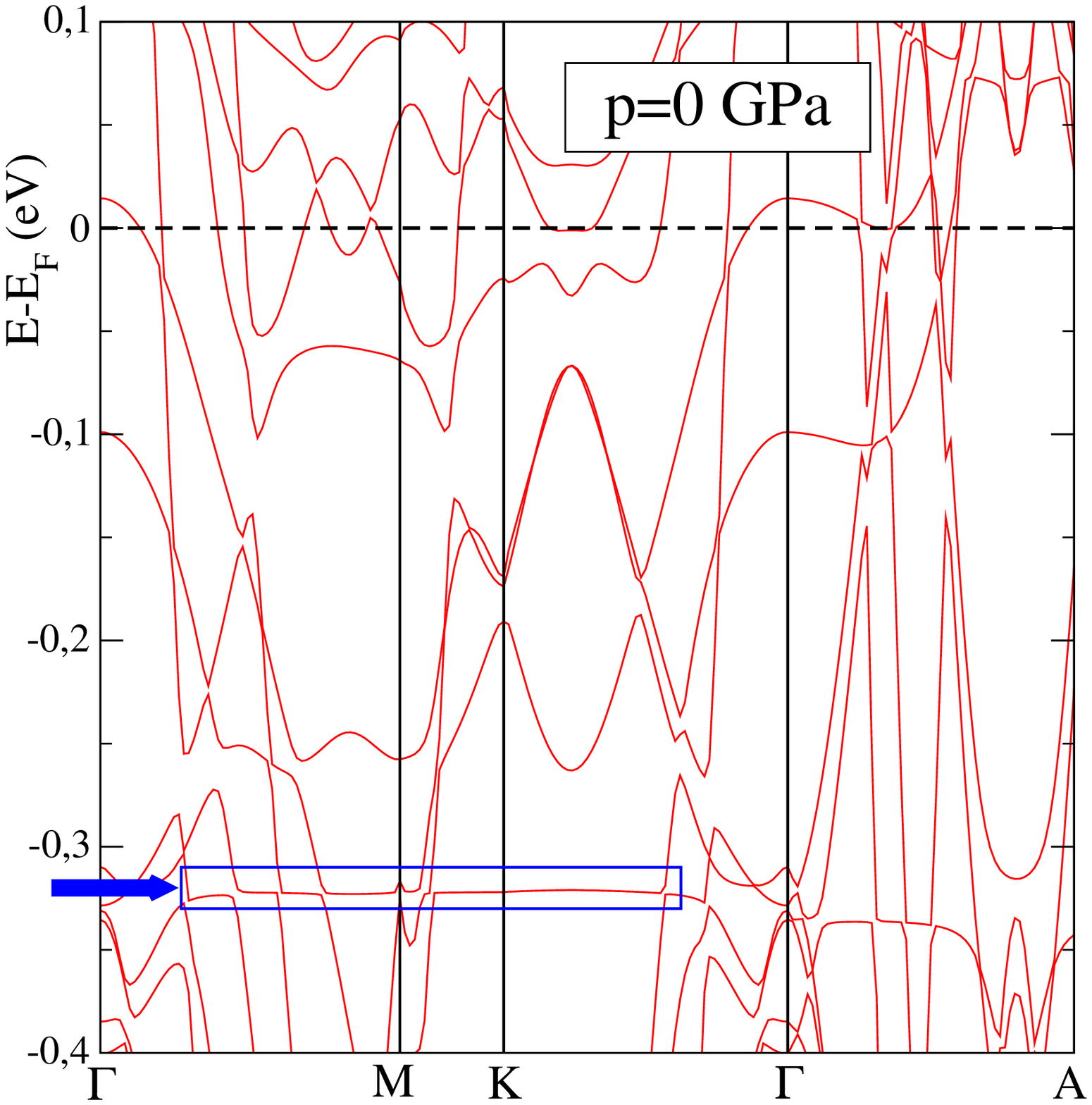}
\includegraphics[scale=0.45, clip=true]{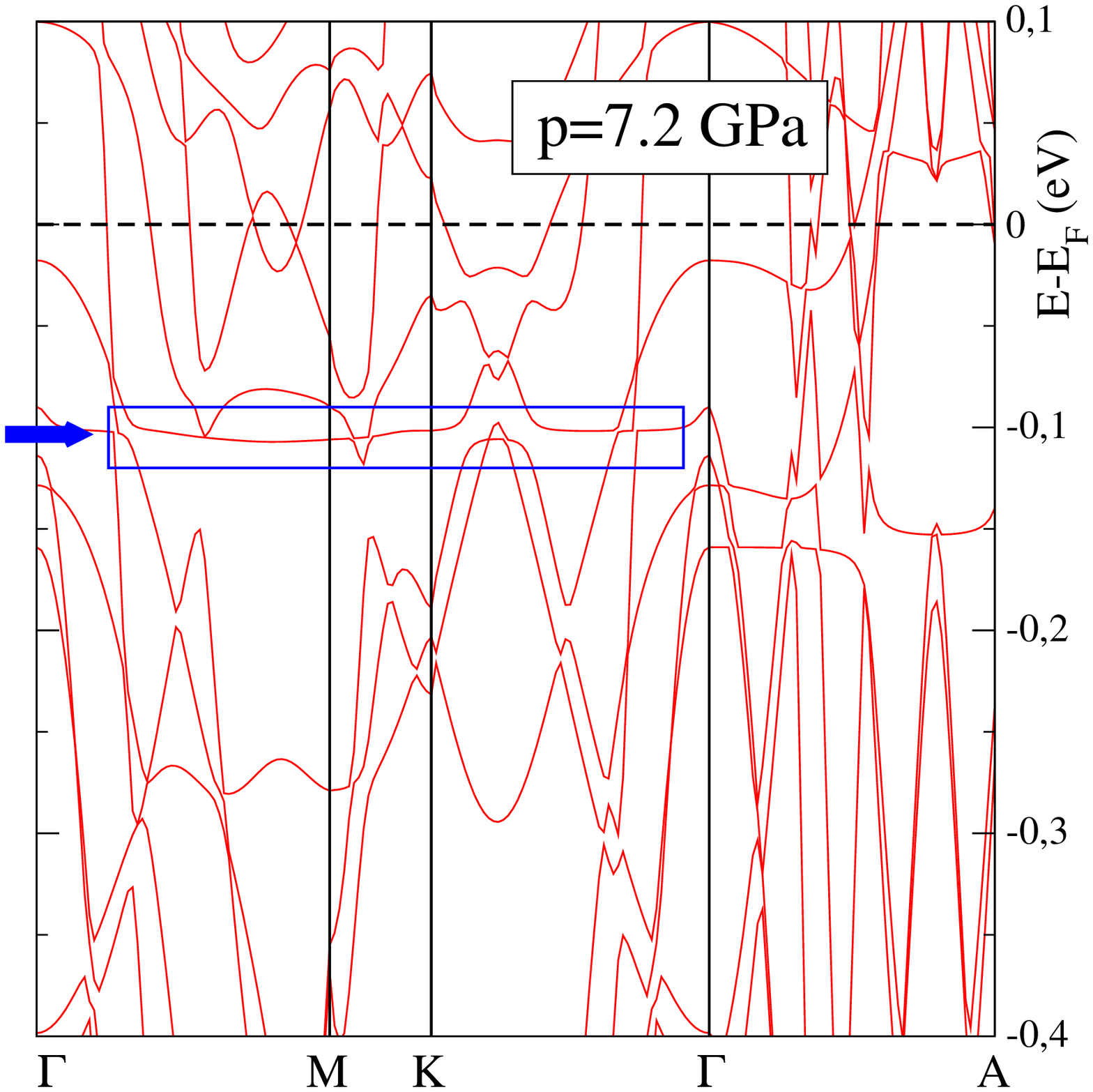}
\end{center}
\caption{Band structure for ambient pressure (left) and 
$p=7.2$ GPa (right) computed within the LSDA including the 
spin-orbit coupling. Computed moments are oriented along the 
$z$-direction. Blue boxes and arrows indicate the position of 
flat bands.}
\label{fig4}
\end{figure*}

The band structure including the SO coupling is presented in Fig~\ref{fig4}. 
At ambient pressure (left panel Fig.~\ref{fig4}) the dispersion less band indicated 
within the blue box is located at about $-0.3eV$ below $E_F$. 
Upon pressure (right panel Fig.~\ref{fig4}) the flat band approaches the Fermi level 
and in the same time a slight departure from the flatness is seen. 
In the absence of spin-orbit coupling we have checked that these flat bands 
reveal their $d_{yz}/d_{zx}$-orbital character in agreement with the previous
results~\cite{ro.ko.06,ko.sc.08}. 
The first order Lifshitz transition was associated with the shifting 
of the dispersion-less band as pressure is increased~\cite{ro.ko.06,ko.sc.08}, 
however electronic correlations were ignored.

\begin{figure}[htp]
\begin{center}
\includegraphics[scale=0.7, clip=true]{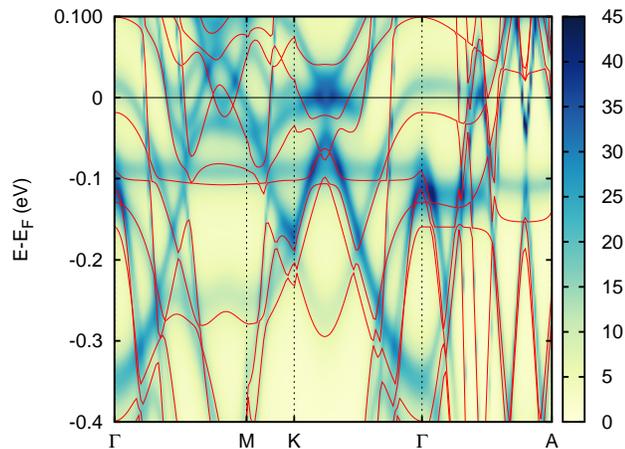}
\end{center}
\caption{LDA+DMFT Bloch spectral function at 7.2 GPa, using $U=2$ eV and $J=0.9$ eV. Red lines correspond to the LSDA band structure.}
\label{bloch}
\end{figure}

In Fig.~\ref{bloch} we compare the LDA band structure of Fig.~\ref{fig4} 
corresponding to the pressure of 7.2 GPa with the LDA+DMFT spectral 
function. Of interest are the bands situated with 0.1eV below $E_F$
where the LDA predict, under pressure, a shifted flat band. This band is further pushed 
towards $E_F$ as a consequence of the negative slope of the real-part of 
the self-energy, however there is no significant change  
in its flatness. Therefore 
electronic correlations and pressure have the same effect in bringing the 
flat-band closer to $E_F$. Consequently, the LDA based prediction
for the threshold pressure for the Lifshitz transition may be decreased 
because of the presence of electronic correlations. Further investigations
are necessary for a quantitative prediction of the threshold value
which goes beyond the scope of the present study.
A more detailed discussion on the 
combined effect of correlation and Lifshitz transition is presented in
Sec.~\ref{sec:disc+concl}.

\subsection{Magnetic moments and orbital polarizations} 
\label{sec:mag_mom}

Our results for the spin $\mu_s$ and orbital $\mu_l$ moments at Co($2c$)/Co($3g$) 
sites are $\mu_s=1.53/1.54 \mu_B$ and $\mu_l=0.22/0.18 \mu_B$ at ambient pressure
and $\mu_s=1.45/1.45 \mu_B$ and $\mu_l=0.18/0.15 \mu_B$ at $7.2$~GPa respectively.
These results can be 
compared with the experimental values given by ND experiment,
see Sec.~\ref{sec:exp_tech}. In the neutron diffraction experiment 
the spin and orbital scattering form factors are not separated, nor 
site resolved, we consider to average the computed moments according to
their different environments, which is sensitive to different pressures. 
The experimentally determined pressure/relative volume dependence of 
cobalt moments is in rather good agreement with the theoretical
data in particular at $V/V_0 > 0.93$. The decrease of 
cobalt moments was previously studied theoretically within the rigid 
band model~\cite{da.ke.91,no.br.92,ya.as.96,st.ri.01}. More recently,
this decrease was associated with a high to a low spin state transition
that takes place at $V/V_0 \approx 0.92$, simultaneously with the 
isomorphic structural change~\cite{ro.ko.06,ko.sc.08}. 

There have been numerous theoretical works on the magnetocrystalline 
anisotropies (MAE) of YCo$_5$
~\cite{st.ri.01,ya.as.96,no.br.92,la.ma.03,no.er.90,zh.ja.14}. 
The MAE values originate from the large orbital magnetic moments of 
Co atoms 
with a slightly larger contribution originating from the $2c$ sites.

\begin{figure}[htp]
\begin{center}
\includegraphics[width=\linewidth, clip=true]{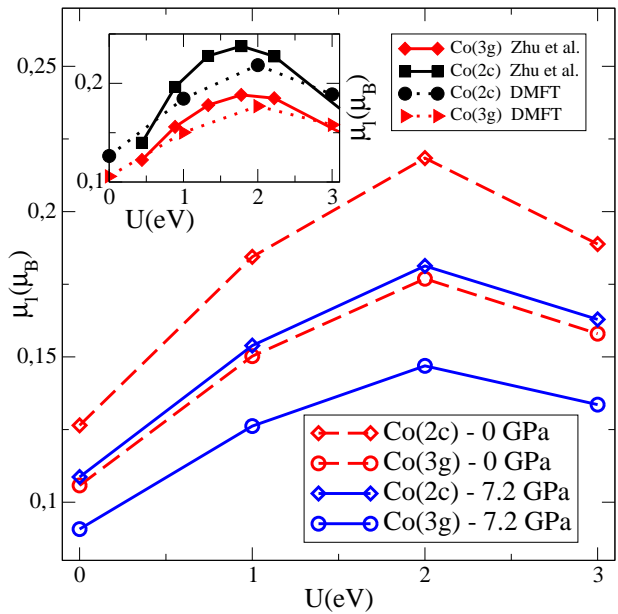}
\end{center}
\caption{Orbital magnetic moments of Co($2c$)/($3g$) atom at ambient 
pressure (red dashed) and at 7.2 GPa (blue solid) as a function of 
the Coulomb parameter $U$, for fixed $J=0.9$ eV. Inset: Ambient pressure
results in comparison with Ref.~\onlinecite{zh.ja.14}.
} 
\label{fig5}
\end{figure}

In Fig.~\ref{fig5} we present the LDA+DMFT results for the computed orbital 
moments as function of the strength of local Coulomb interaction $U$ and for a
fixed value of the exchange parameter $J$. We preliminary checked that
changes in the orbital moment are very small with respect to $J$, so we
varied $U$ within the reasonable range of values for transition metal 
elements, while keeping $J=0.9$~eV fixed. The atom resolved Co($2c$) and Co($3g$) orbital moments reach 
maximum values for $U$ about 2~eV. 
Our results for the orbital moments are consistent with previous 
X-ray magnetic circular experiments~\cite{zh.ja.14},
inelastic spin flip neutron scattering experiments~\cite{he.ri.75}
and calculations including the orbital polarization scheme~\cite{st.ri.01,no.br.92} 
or the recent LDA+DMFT~\cite{zh.ja.14}.

The magnetic properties of Y are induced by the hybridization with
neighboring Co atoms in the basal-($ab$) plane. 
We observe that upon increasing pressure the 
Y$4d$-Co$3d$ hybridization and consequently the negative polarization 
induced on Y$4d$ bands increases. Its magnitude which we 
denote by M$4d$, depends on the number of cobalt atoms situated in 
the first coordination shell to an Y atom ($z_i$) and their magnetic 
moments $M_{Co_i}$. 
Note that for YCo$_5$ the ratio 
$\vert M_{4d} \vert/\sum_i z_i M_{Co_i} = 1.2 \cdot 10^{-2}$ has nearly
the same value as evidenced in RM$_5$ compounds (where M = Co, Ni and 
R = heavy rare-earth), for which a value $1.3 \cdot 10^{-2}$ 
was obtained~\cite{bu.ch.11}. Therefore the d-bands of the transition metal 
elements not only mediate the dominant $4f$-magnetism 
of heavy rare-earths in the RM$_5$ compounds, but also polarize the 
R$5d$ band of rare-earth atoms. Our estimation for the ratio
$\vert M_{4d} \vert/\sum_i z_i M_{Co_i}$ in YCo$_5$, show that also 
in the absence of dominant $4f$-magnetism this effect is still present, 
with a similar intensity.

\section{Discussions and Conclusion}
\label{sec:disc+concl}

According to Koudela \emph{et al.}~\cite{ko.sc.08}, the first order 
structural transition under pressure can be ascribed to a Lifshitz
transition. However, in this scenario the many-body electronic correlation effects
are completely disregarded. On the other hand electronic correlations 
are essential to capture the considerable orbital polarization in this compound.
A specific question which appears is the interplay of the electronic
correlation and the Lifshitz transition. In this section we discuss 
the behavior of the flat-band (related to the Lifshitz transition) 
and the energy dependence of the imaginary part of the self-energy (related
to many-body effects) of YCo$_5$ under pressure.

\begin{figure}[htp]
\begin{center}
\includegraphics[width=\linewidth, clip=true]{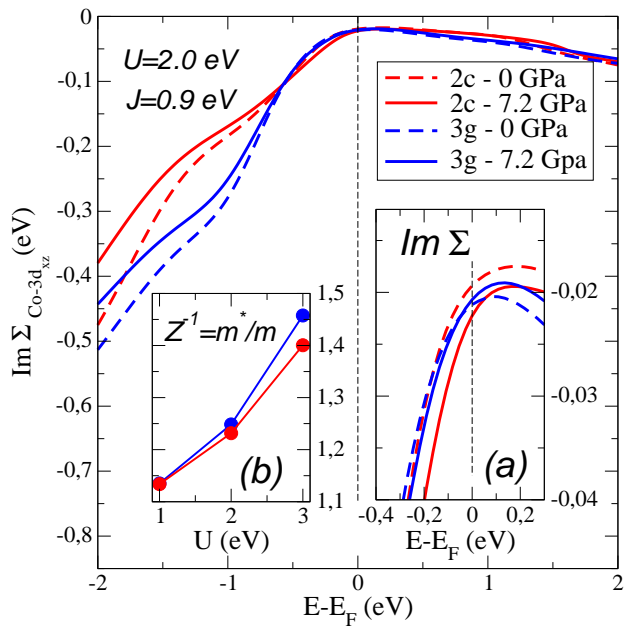}
\end{center}
\caption{Self-energies for the $d_{xz}$-orbital at ambient pressure 
(dashed)
and at $p=7.2$ GPa (solid), for Co atoms at ($2c$/$3g$) site with red/blue
lines. Inset (a)
a reduced energy window around $E_F$. Inset (b) Effective mass
renormalization as function of $U$ at ambient pressure.}
\label{fig6}
\end{figure}

In Fig.~\ref{fig6} we show the imaginary part of the self-energy for the 
Co($2c$) and Co($3g$) atoms for the selected $d_{xz}$ orbital contributing to the flat 
bands. All the other orbitals have a qualitatively similar energy-dependence of the imaginary part of the self-energy. 
The results for the ambient pressure are presented with dashed lines, while 
continuous lines are used for the results at 7.2~GPa. The right inset presents the 
$Im~\Sigma_{xz}(E)$ in the energy window $[-0.4eV, 0.3eV]$ around $E_F$.
At ambient pressure, the dispersion-less band is situated at larger distance from 
$E_F$, therefore the system is away from the Lifshitz transition. 
The self-energy follow the Fermi-liquid behavior, i.e. a parabolic energy dependence
$Im~\Sigma_{xz}(E) \propto (E-E_F)^2$. 
At 7.2~GPa the flat-band approaches the Fermi level and a slight departure
from the flatness can be seen (Fig.~\ref{fig4}). Electronic correlations
push the dispersion less band even closer to the $E_F$.

Local on-site correlation effects are dominant in the scattering rates related 
to electron-electron interaction and can be studied in the framework of LDA+DMFT.
The quasiparticle scattering rate $\Gamma_{ee}$ of the charge carriers can be 
computed from the imaginary part of the self-energy, according to the formula: 
$\Gamma_{ee} = Z \cdot \mathrm{Im} \Sigma_{xz}(E)$, where $Z^{-1} = (m^*/m_{LDA})$ 
is the renormalization factor which in case of electronic correlations reduces the 
step of momentum density at $k_F$. In the left inset of Fig.~\ref{fig6} the results
for the mass enhancement as a function of the strength of Coulomb interaction $U$
is presented. This is computed from the self-energy according to: 
$m^*/m_{LDA}=[1-\partial \mathrm{Im} \Sigma / \partial \omega_n|_{i\omega_n \rightarrow 0}]^{-1}$. The effective mass
monotonically increases with $U$ similar to the previous results~\cite{zh.ja.14}, 
implying a monotonic decrease of $Z$.
This term contributes as a multiplicative constant to the 
$\Sigma_{xz}(E)$ consequently, the scattering rate follows also a Fermi-liquid behavior.
Resistivity under pressure was shown to follow
a similar behavior in RCo$_5$-compounds~\cite{ho.ha.95}.

In summary, the present paper discusses the 
properties of YCo$_5$ near a structural transition induced by pressure.
Our neutron diffraction experiment and the corresponding theoretical 
modeling within LDA+DMFT, is performed up to a pressure of 7.2 GPa. 
Although at this pressure the system does not reach the possible Lifshitz transition
proposed previously~\cite{ro.ko.06,ko.sc.08}, we study the behavior of the 
bands, density of states and magnetic moments.
Our many-body calculations reveal that the imaginary part of the 
self-energy follows a parabolic energy dependence approaching 
the Lifshitz transition. However, for stronger Coulomb $U$ parameters
the slope of the real part of the self-energy increases, and accelerate
the proximity to the Lifshitz transition. The increase in the effective 
mass acts as a multiplicative constant to the imaginary part of the self-energy,
therefore the scattering rate would also follow the Fermi-liquid behavior.
Finally, according to our results the physical properties of YCo$_5$ under 
pressure could be described supplementing the Lifshitz
scenario~\cite{ro.ko.06,ko.sc.08,oc.ry.15} with  
strong local electron-electron interaction.

\section*{Acknowledgment}
A.\"O and L.C. acknowledges financial support offered by the Augsburg 
Center for Innovative Technologies, and by the Deutsche
Forschungsgemeinschaft (through the project TRR 80/F6).

\bibliography{main}

\begin{thebibliography}{42}%
\makeatletter
\providecommand \@ifxundefined [1]{%
 \@ifx{#1\undefined}
}%
\providecommand \@ifnum [1]{%
 \ifnum #1\expandafter \@firstoftwo
 \else \expandafter \@secondoftwo
 \fi
}%
\providecommand \@ifx [1]{%
 \ifx #1\expandafter \@firstoftwo
 \else \expandafter \@secondoftwo
 \fi
}%
\providecommand \natexlab [1]{#1}%
\providecommand \enquote  [1]{``#1''}%
\providecommand \bibnamefont  [1]{#1}%
\providecommand \bibfnamefont [1]{#1}%
\providecommand \citenamefont [1]{#1}%
\providecommand \href@noop [0]{\@secondoftwo}%
\providecommand \href [0]{\begingroup \@sanitize@url \@href}%
\providecommand \@href[1]{\@@startlink{#1}\@@href}%
\providecommand \@@href[1]{\endgroup#1\@@endlink}%
\providecommand \@sanitize@url [0]{\catcode `\\12\catcode `\$12\catcode
  `\&12\catcode `\#12\catcode `\^12\catcode `\_12\catcode `\%12\relax}%
\providecommand \@@startlink[1]{}%
\providecommand \@@endlink[0]{}%
\providecommand \url  [0]{\begingroup\@sanitize@url \@url }%
\providecommand \@url [1]{\endgroup\@href {#1}{\urlprefix }}%
\providecommand \urlprefix  [0]{URL }%
\providecommand \Eprint [0]{\href }%
\providecommand \doibase [0]{http://dx.doi.org/}%
\providecommand \selectlanguage [0]{\@gobble}%
\providecommand \bibinfo  [0]{\@secondoftwo}%
\providecommand \bibfield  [0]{\@secondoftwo}%
\providecommand \translation [1]{[#1]}%
\providecommand \BibitemOpen [0]{}%
\providecommand \bibitemStop [0]{}%
\providecommand \bibitemNoStop [0]{.\EOS\space}%
\providecommand \EOS [0]{\spacefactor3000\relax}%
\providecommand \BibitemShut  [1]{\csname bibitem#1\endcsname}%
\let\auto@bib@innerbib\@empty
\bibitem [{\citenamefont {Burzo}\ \emph {et~al.}(1990)\citenamefont {Burzo},
  \citenamefont {Chelkowski},\ and\ \citenamefont {Kirchmayr}}]{bu.ch.90}%
  \BibitemOpen
  \bibfield  {author} {\bibinfo {author} {\bibfnamefont {E.}~\bibnamefont
  {Burzo}}, \bibinfo {author} {\bibfnamefont {A.}~\bibnamefont {Chelkowski}}, \
  and\ \bibinfo {author} {\bibfnamefont {H.~R.}\ \bibnamefont {Kirchmayr}},\
  }in\ \href@noop {} {\emph {\bibinfo {booktitle} {Landolt B\"ornstein
  Handbook}}},\ Vol.\ \bibinfo {volume} {19d2}\ (\bibinfo  {publisher}
  {Springer Verlag},\ \bibinfo {address} {Heidelberg},\ \bibinfo {year}
  {1990})\ p.\ \bibinfo {pages} {130}\BibitemShut {NoStop}%
\bibitem [{\citenamefont {Rosner}\ \emph {et~al.}(2006)\citenamefont {Rosner},
  \citenamefont {Koudela}, \citenamefont {Schwarz}, \citenamefont {Handstein},
  \citenamefont {Hanfland}, \citenamefont {Opahle}, \citenamefont {Koepernik},
  \citenamefont {Kuz'min}, \citenamefont {M\"uller}, \citenamefont {Mydosh},\
  and\ \citenamefont {Richter}}]{ro.ko.06}%
  \BibitemOpen
  \bibfield  {author} {\bibinfo {author} {\bibfnamefont {H.}~\bibnamefont
  {Rosner}}, \bibinfo {author} {\bibfnamefont {D.}~\bibnamefont {Koudela}},
  \bibinfo {author} {\bibfnamefont {U.}~\bibnamefont {Schwarz}}, \bibinfo
  {author} {\bibfnamefont {A.}~\bibnamefont {Handstein}}, \bibinfo {author}
  {\bibfnamefont {M.}~\bibnamefont {Hanfland}}, \bibinfo {author}
  {\bibfnamefont {I.}~\bibnamefont {Opahle}}, \bibinfo {author} {\bibfnamefont
  {K.}~\bibnamefont {Koepernik}}, \bibinfo {author} {\bibfnamefont {M.~D.}\
  \bibnamefont {Kuz'min}}, \bibinfo {author} {\bibfnamefont {K.~H.}\
  \bibnamefont {M\"uller}}, \bibinfo {author} {\bibfnamefont {J.~A.}\
  \bibnamefont {Mydosh}}, \ and\ \bibinfo {author} {\bibfnamefont
  {M.}~\bibnamefont {Richter}},\ }\href {\doibase 10.1038/nphys341} {\bibfield
  {journal} {\bibinfo  {journal} {Nature Physics}\ }\textbf {\bibinfo {volume}
  {2}},\ \bibinfo {pages} {469} (\bibinfo {year} {2006})}\BibitemShut {NoStop}%
\bibitem [{\citenamefont {Koudela}\ \emph {et~al.}(2008)\citenamefont
  {Koudela}, \citenamefont {Schwarz}, \citenamefont {Rosner}, \citenamefont
  {Burkhardt}, \citenamefont {Handstein}, \citenamefont {Hanfland},
  \citenamefont {Kuz'min}, \citenamefont {Opahle}, \citenamefont {Koepernik},
  \citenamefont {M\"uller},\ and\ \citenamefont {Richter}}]{ko.sc.08}%
  \BibitemOpen
  \bibfield  {author} {\bibinfo {author} {\bibfnamefont {D.}~\bibnamefont
  {Koudela}}, \bibinfo {author} {\bibfnamefont {U.}~\bibnamefont {Schwarz}},
  \bibinfo {author} {\bibfnamefont {H.}~\bibnamefont {Rosner}}, \bibinfo
  {author} {\bibfnamefont {U.}~\bibnamefont {Burkhardt}}, \bibinfo {author}
  {\bibfnamefont {A.}~\bibnamefont {Handstein}}, \bibinfo {author}
  {\bibfnamefont {M.}~\bibnamefont {Hanfland}}, \bibinfo {author}
  {\bibfnamefont {M.~D.}\ \bibnamefont {Kuz'min}}, \bibinfo {author}
  {\bibfnamefont {I.}~\bibnamefont {Opahle}}, \bibinfo {author} {\bibfnamefont
  {K.}~\bibnamefont {Koepernik}}, \bibinfo {author} {\bibfnamefont {K.-H.}\
  \bibnamefont {M\"uller}}, \ and\ \bibinfo {author} {\bibfnamefont
  {M.}~\bibnamefont {Richter}},\ }\href {\doibase 10.1103/PhysRevB.77.024411}
  {\bibfield  {journal} {\bibinfo  {journal} {Phys. Rev. B}\ }\textbf {\bibinfo
  {volume} {77}},\ \bibinfo {pages} {024411} (\bibinfo {year}
  {2008})}\BibitemShut {NoStop}%
\bibitem [{\citenamefont {Burzo}\ and\ \citenamefont {Vlaic}(2013)}]{bu.vl.13}%
  \BibitemOpen
  \bibfield  {author} {\bibinfo {author} {\bibfnamefont {E.}~\bibnamefont
  {Burzo}}\ and\ \bibinfo {author} {\bibfnamefont {P.}~\bibnamefont {Vlaic}},\
  }\href@noop {} {\bibfield  {journal} {\bibinfo  {journal} {AIP Conf. Proc.}\
  }\textbf {\bibinfo {volume} {1564}},\ \bibinfo {pages} {103} (\bibinfo {year}
  {2013})}\BibitemShut {NoStop}%
\bibitem [{\citenamefont {Ochi}\ \emph
  {et~al.}(2015{\natexlab{a}})\citenamefont {Ochi}, \citenamefont {Arita},
  \citenamefont {Matsumoto}, \citenamefont {Kino},\ and\ \citenamefont
  {Miyake}}]{ma.ar.15}%
  \BibitemOpen
  \bibfield  {author} {\bibinfo {author} {\bibfnamefont {M.}~\bibnamefont
  {Ochi}}, \bibinfo {author} {\bibfnamefont {R.}~\bibnamefont {Arita}},
  \bibinfo {author} {\bibfnamefont {M.}~\bibnamefont {Matsumoto}}, \bibinfo
  {author} {\bibfnamefont {H.}~\bibnamefont {Kino}}, \ and\ \bibinfo {author}
  {\bibfnamefont {T.}~\bibnamefont {Miyake}},\ }\href {\doibase
  10.1103/PhysRevB.91.165137} {\bibfield  {journal} {\bibinfo  {journal} {Phys.
  Rev. B}\ }\textbf {\bibinfo {volume} {91}},\ \bibinfo {pages} {165137}
  (\bibinfo {year} {2015}{\natexlab{a}})}\BibitemShut {NoStop}%
\bibitem [{\citenamefont {Lifshitz}(1960)}]{lifs.60}%
  \BibitemOpen
  \bibfield  {author} {\bibinfo {author} {\bibfnamefont {I.~M.}\ \bibnamefont
  {Lifshitz}},\ }\href@noop {} {\bibfield  {journal} {\bibinfo  {journal} {Sov.
  Phys. JETP}\ }\textbf {\bibinfo {volume} {11}},\ \bibinfo {pages} {1130}
  (\bibinfo {year} {1960})}\BibitemShut {NoStop}%
\bibitem [{\citenamefont {Kohn}(1999)}]{kohn.99}%
  \BibitemOpen
  \bibfield  {author} {\bibinfo {author} {\bibfnamefont {W.}~\bibnamefont
  {Kohn}},\ }\href@noop {} {\bibfield  {journal} {\bibinfo  {journal} {Rev.
  Mod. Phys.}\ }\textbf {\bibinfo {volume} {71}},\ \bibinfo {pages} {1253}
  (\bibinfo {year} {1999})}\BibitemShut {NoStop}%
\bibitem [{\citenamefont {Metzner}\ and\ \citenamefont
  {Vollhardt}(1989)}]{me.vo.89}%
  \BibitemOpen
  \bibfield  {author} {\bibinfo {author} {\bibfnamefont {W.}~\bibnamefont
  {Metzner}}\ and\ \bibinfo {author} {\bibfnamefont {D.}~\bibnamefont
  {Vollhardt}},\ }\href@noop {} {\bibfield  {journal} {\bibinfo  {journal}
  {Phys. Rev. Lett.}\ }\textbf {\bibinfo {volume} {62}},\ \bibinfo {pages}
  {324} (\bibinfo {year} {1989})}\BibitemShut {NoStop}%
\bibitem [{\citenamefont {Kotliar}\ and\ \citenamefont
  {Vollhardt}(2004)}]{ko.vo.04}%
  \BibitemOpen
  \bibfield  {author} {\bibinfo {author} {\bibfnamefont {G.}~\bibnamefont
  {Kotliar}}\ and\ \bibinfo {author} {\bibfnamefont {D.}~\bibnamefont
  {Vollhardt}},\ }\href@noop {} {\bibfield  {journal} {\bibinfo  {journal}
  {Physics Today}\ }\textbf {\bibinfo {volume} {57}},\ \bibinfo {pages} {53}
  (\bibinfo {year} {2004})}\BibitemShut {NoStop}%
\bibitem [{\citenamefont {Kotliar}\ \emph {et~al.}(2006)\citenamefont
  {Kotliar}, \citenamefont {Savrasov}, \citenamefont {Haule}, \citenamefont
  {Oudovenko}, \citenamefont {Parcollet},\ and\ \citenamefont
  {Marianetti}}]{ko.sa.06}%
  \BibitemOpen
  \bibfield  {author} {\bibinfo {author} {\bibfnamefont {G.}~\bibnamefont
  {Kotliar}}, \bibinfo {author} {\bibfnamefont {S.~Y.}\ \bibnamefont
  {Savrasov}}, \bibinfo {author} {\bibfnamefont {K.}~\bibnamefont {Haule}},
  \bibinfo {author} {\bibfnamefont {V.~S.}\ \bibnamefont {Oudovenko}}, \bibinfo
  {author} {\bibfnamefont {O.}~\bibnamefont {Parcollet}}, \ and\ \bibinfo
  {author} {\bibfnamefont {C.~A.}\ \bibnamefont {Marianetti}},\ }\href@noop {}
  {\bibfield  {journal} {\bibinfo  {journal} {Rev. Mod. Phys.}\ }\textbf
  {\bibinfo {volume} {78}},\ \bibinfo {pages} {865} (\bibinfo {year}
  {2006})}\BibitemShut {NoStop}%
\bibitem [{\citenamefont {Held}(2007)}]{held.07}%
  \BibitemOpen
  \bibfield  {author} {\bibinfo {author} {\bibfnamefont {K.}~\bibnamefont
  {Held}},\ }\href@noop {} {\bibfield  {journal} {\bibinfo  {journal} {Adv.
  Phys.}\ }\textbf {\bibinfo {volume} {56}},\ \bibinfo {pages} {829} (\bibinfo
  {year} {2007})}\BibitemShut {NoStop}%
\bibitem [{\citenamefont {Zhu}\ \emph {et~al.}(2014)\citenamefont {Zhu},
  \citenamefont {Janoschek}, \citenamefont {Rosenberg}, \citenamefont
  {Ronning}, \citenamefont {Thompson}, \citenamefont {Torrez}, \citenamefont
  {Bauer},\ and\ \citenamefont {Batista}}]{zh.ja.14}%
  \BibitemOpen
  \bibfield  {author} {\bibinfo {author} {\bibfnamefont {J.-X.}\ \bibnamefont
  {Zhu}}, \bibinfo {author} {\bibfnamefont {M.}~\bibnamefont {Janoschek}},
  \bibinfo {author} {\bibfnamefont {R.}~\bibnamefont {Rosenberg}}, \bibinfo
  {author} {\bibfnamefont {F.}~\bibnamefont {Ronning}}, \bibinfo {author}
  {\bibfnamefont {J.~D.}\ \bibnamefont {Thompson}}, \bibinfo {author}
  {\bibfnamefont {M.~A.}\ \bibnamefont {Torrez}}, \bibinfo {author}
  {\bibfnamefont {E.~D.}\ \bibnamefont {Bauer}}, \ and\ \bibinfo {author}
  {\bibfnamefont {C.~D.}\ \bibnamefont {Batista}},\ }\href {\doibase
  10.1103/PhysRevX.4.021027} {\bibfield  {journal} {\bibinfo  {journal} {Phys.
  Rev. X}\ }\textbf {\bibinfo {volume} {4}},\ \bibinfo {pages} {021027}
  (\bibinfo {year} {2014})}\BibitemShut {NoStop}%
\bibitem [{\citenamefont {Aksenov}\ \emph {et~al.}(1999)\citenamefont
  {Aksenov}, \citenamefont {Balagurov}, \citenamefont {Glazkov}, \citenamefont
  {Kozlenko}, \citenamefont {Naumov}, \citenamefont {Savenko}, \citenamefont
  {Sheptyakov}, \citenamefont {Somenkov}, \citenamefont {Bulkin}, \citenamefont
  {Kudryashev},\ and\ \citenamefont {Trounov}}]{ak.ba.99}%
  \BibitemOpen
  \bibfield  {author} {\bibinfo {author} {\bibfnamefont {V.}~\bibnamefont
  {Aksenov}}, \bibinfo {author} {\bibfnamefont {A.}~\bibnamefont {Balagurov}},
  \bibinfo {author} {\bibfnamefont {V.}~\bibnamefont {Glazkov}}, \bibinfo
  {author} {\bibfnamefont {D.}~\bibnamefont {Kozlenko}}, \bibinfo {author}
  {\bibfnamefont {I.}~\bibnamefont {Naumov}}, \bibinfo {author} {\bibfnamefont
  {B.}~\bibnamefont {Savenko}}, \bibinfo {author} {\bibfnamefont
  {D.}~\bibnamefont {Sheptyakov}}, \bibinfo {author} {\bibfnamefont
  {V.}~\bibnamefont {Somenkov}}, \bibinfo {author} {\bibfnamefont
  {A.}~\bibnamefont {Bulkin}}, \bibinfo {author} {\bibfnamefont
  {V.}~\bibnamefont {Kudryashev}}, \ and\ \bibinfo {author} {\bibfnamefont
  {V.}~\bibnamefont {Trounov}},\ }\href@noop {} {\bibfield  {journal} {\bibinfo
   {journal} {Physica B}\ }\textbf {\bibinfo {volume} {265}},\ \bibinfo {pages}
  {258} (\bibinfo {year} {1999})}\BibitemShut {NoStop}%
\bibitem [{\citenamefont {Glazkov}\ and\ \citenamefont
  {Goncharenko}(1991)}]{gl.go.91}%
  \BibitemOpen
  \bibfield  {author} {\bibinfo {author} {\bibfnamefont {V.~P.}\ \bibnamefont
  {Glazkov}}\ and\ \bibinfo {author} {\bibfnamefont {L.}~\bibnamefont
  {Goncharenko}},\ }\href@noop {} {\bibfield  {journal} {\bibinfo  {journal}
  {Fiz. Tekh. Vys. Davlenii}\ }\textbf {\bibinfo {volume} {1}},\ \bibinfo
  {pages} {56} (\bibinfo {year} {1991})}\BibitemShut {NoStop}%
\bibitem [{\citenamefont {Zlokazov}\ and\ \citenamefont
  {Chernyshev}(1992)}]{zl.ch.91}%
  \BibitemOpen
  \bibfield  {author} {\bibinfo {author} {\bibfnamefont {V.~B.}\ \bibnamefont
  {Zlokazov}}\ and\ \bibinfo {author} {\bibfnamefont {V.~U.}\ \bibnamefont
  {Chernyshev}},\ }\href@noop {} {\bibfield  {journal} {\bibinfo  {journal} {J.
  Appl. Crystallogr.}\ }\textbf {\bibinfo {volume} {25}},\ \bibinfo {pages}
  {447} (\bibinfo {year} {1992})}\BibitemShut {NoStop}%
\bibitem [{\citenamefont {Rodriguez-Carvajal}(1993)}]{rodr.93}%
  \BibitemOpen
  \bibfield  {author} {\bibinfo {author} {\bibfnamefont {J.}~\bibnamefont
  {Rodriguez-Carvajal}},\ }\href@noop {} {\bibfield  {journal} {\bibinfo
  {journal} {Physica B}\ }\textbf {\bibinfo {volume} {192}},\ \bibinfo {pages}
  {55} (\bibinfo {year} {1993})}\BibitemShut {NoStop}%
\bibitem [{\citenamefont {Patrick}\ \emph {et~al.}(2017)\citenamefont
  {Patrick}, \citenamefont {Kumar}, \citenamefont {Balakrishnan}, \citenamefont
  {Edwards}, \citenamefont {Lees}, \citenamefont {Mendive-Tapia}, \citenamefont
  {Petit},\ and\ \citenamefont {Staunton}}]{ch.sa.17}%
  \BibitemOpen
  \bibfield  {author} {\bibinfo {author} {\bibfnamefont {C.~E.}\ \bibnamefont
  {Patrick}}, \bibinfo {author} {\bibfnamefont {S.}~\bibnamefont {Kumar}},
  \bibinfo {author} {\bibfnamefont {G.}~\bibnamefont {Balakrishnan}}, \bibinfo
  {author} {\bibfnamefont {R.~S.}\ \bibnamefont {Edwards}}, \bibinfo {author}
  {\bibfnamefont {M.~R.}\ \bibnamefont {Lees}}, \bibinfo {author}
  {\bibfnamefont {E.}~\bibnamefont {Mendive-Tapia}}, \bibinfo {author}
  {\bibfnamefont {L.}~\bibnamefont {Petit}}, \ and\ \bibinfo {author}
  {\bibfnamefont {J.~B.}\ \bibnamefont {Staunton}},\ }\href {\doibase
  10.1103/PhysRevMaterials.1.024411} {\bibfield  {journal} {\bibinfo  {journal}
  {Phys. Rev. Materials}\ }\textbf {\bibinfo {volume} {1}},\ \bibinfo {pages}
  {024411} (\bibinfo {year} {2017})}\BibitemShut {NoStop}%
\bibitem [{\citenamefont {Buschow}(1977)}]{busc.77}%
  \BibitemOpen
  \bibfield  {author} {\bibinfo {author} {\bibfnamefont {K.~H.~J.}\
  \bibnamefont {Buschow}},\ }\href@noop {} {\bibfield  {journal} {\bibinfo
  {journal} {Rep. Prog. Phys}\ }\textbf {\bibinfo {volume} {40}},\ \bibinfo
  {pages} {1179} (\bibinfo {year} {1977})}\BibitemShut {NoStop}%
\bibitem [{\citenamefont {Daalderop}\ \emph {et~al.}(1996)\citenamefont
  {Daalderop}, \citenamefont {Kelly},\ and\ \citenamefont
  {Schuurmans}}]{da.ke.96}%
  \BibitemOpen
  \bibfield  {author} {\bibinfo {author} {\bibfnamefont {G.~H.~O.}\
  \bibnamefont {Daalderop}}, \bibinfo {author} {\bibfnamefont {P.~J.}\
  \bibnamefont {Kelly}}, \ and\ \bibinfo {author} {\bibfnamefont {M.~F.~H.}\
  \bibnamefont {Schuurmans}},\ }\href {\doibase 10.1103/PhysRevB.53.14415}
  {\bibfield  {journal} {\bibinfo  {journal} {Phys. Rev. B}\ }\textbf {\bibinfo
  {volume} {53}},\ \bibinfo {pages} {14415} (\bibinfo {year}
  {1996})}\BibitemShut {NoStop}%
\bibitem [{\citenamefont {Deportes}\ \emph {et~al.}(1976)\citenamefont
  {Deportes}, \citenamefont {Givord}, \citenamefont {Schweizer},\ and\
  \citenamefont {Tasset}}]{de.gi.76}%
  \BibitemOpen
  \bibfield  {author} {\bibinfo {author} {\bibfnamefont {J.}~\bibnamefont
  {Deportes}}, \bibinfo {author} {\bibfnamefont {D.}~\bibnamefont {Givord}},
  \bibinfo {author} {\bibfnamefont {J.}~\bibnamefont {Schweizer}}, \ and\
  \bibinfo {author} {\bibfnamefont {F.}~\bibnamefont {Tasset}},\ }\href@noop {}
  {\bibfield  {journal} {\bibinfo  {journal} {IEEE Trans. Magn. MAG}\ }\textbf
  {\bibinfo {volume} {12}},\ \bibinfo {pages} {1000} (\bibinfo {year}
  {1976})}\BibitemShut {NoStop}%
\bibitem [{\citenamefont {Streever}(1978)}]{stre.78}%
  \BibitemOpen
  \bibfield  {author} {\bibinfo {author} {\bibfnamefont {R.~L.}\ \bibnamefont
  {Streever}},\ }\href@noop {} {\bibfield  {journal} {\bibinfo  {journal}
  {Phys. Letters}\ }\textbf {\bibinfo {volume} {63A}},\ \bibinfo {pages} {360}
  (\bibinfo {year} {1978})}\BibitemShut {NoStop}%
\bibitem [{\citenamefont {Streever}(1979)}]{stre.79}%
  \BibitemOpen
  \bibfield  {author} {\bibinfo {author} {\bibfnamefont {R.~L.}\ \bibnamefont
  {Streever}},\ }\href {\doibase 10.1103/PhysRevB.19.2704} {\bibfield
  {journal} {\bibinfo  {journal} {Phys. Rev. B}\ }\textbf {\bibinfo {volume}
  {19}},\ \bibinfo {pages} {2704} (\bibinfo {year} {1979})}\BibitemShut
  {NoStop}%
\bibitem [{\citenamefont {Daalderop}\ \emph {et~al.}(1991)\citenamefont
  {Daalderop}, \citenamefont {Kelly},\ and\ \citenamefont
  {Schuurmans}}]{da.ke.91}%
  \BibitemOpen
  \bibfield  {author} {\bibinfo {author} {\bibfnamefont {G.~H.~O.}\
  \bibnamefont {Daalderop}}, \bibinfo {author} {\bibfnamefont {P.~J.}\
  \bibnamefont {Kelly}}, \ and\ \bibinfo {author} {\bibfnamefont {M.~F.~H.}\
  \bibnamefont {Schuurmans}},\ }\href {\doibase 10.1103/PhysRevB.44.12054}
  {\bibfield  {journal} {\bibinfo  {journal} {Phys. Rev. B}\ }\textbf {\bibinfo
  {volume} {44}},\ \bibinfo {pages} {12054} (\bibinfo {year}
  {1991})}\BibitemShut {NoStop}%
\bibitem [{\citenamefont {Nordstr\"om}\ \emph {et~al.}(1992)\citenamefont
  {Nordstr\"om}, \citenamefont {Brooks},\ and\ \citenamefont
  {Johansson}}]{no.br.92}%
  \BibitemOpen
  \bibfield  {author} {\bibinfo {author} {\bibfnamefont {L.}~\bibnamefont
  {Nordstr\"om}}, \bibinfo {author} {\bibfnamefont {M.~S.~S.}\ \bibnamefont
  {Brooks}}, \ and\ \bibinfo {author} {\bibfnamefont {B.}~\bibnamefont
  {Johansson}},\ }\href@noop {} {\bibfield  {journal} {\bibinfo  {journal} {J.
  Phys: Condens. Matter}\ }\textbf {\bibinfo {volume} {4}},\ \bibinfo {pages}
  {3261} (\bibinfo {year} {1992})}\BibitemShut {NoStop}%
\bibitem [{\citenamefont {Yamaguchi}\ and\ \citenamefont
  {Asano}(1996)}]{ya.as.96}%
  \BibitemOpen
  \bibfield  {author} {\bibinfo {author} {\bibfnamefont {M.}~\bibnamefont
  {Yamaguchi}}\ and\ \bibinfo {author} {\bibfnamefont {S.}~\bibnamefont
  {Asano}},\ }\href@noop {} {\bibfield  {journal} {\bibinfo  {journal} {J.
  Appl. Phys.}\ }\textbf {\bibinfo {volume} {79}},\ \bibinfo {pages} {5952}
  (\bibinfo {year} {1996})}\BibitemShut {NoStop}%
\bibitem [{\citenamefont {Jansen}(1999)}]{jans.99}%
  \BibitemOpen
  \bibfield  {author} {\bibinfo {author} {\bibfnamefont {H.~J.~F.}\
  \bibnamefont {Jansen}},\ }\href {\doibase 10.1103/PhysRevB.59.4699}
  {\bibfield  {journal} {\bibinfo  {journal} {Phys. Rev. B}\ }\textbf {\bibinfo
  {volume} {59}},\ \bibinfo {pages} {4699} (\bibinfo {year}
  {1999})}\BibitemShut {NoStop}%
\bibitem [{\citenamefont {Yamada}\ \emph {et~al.}(1991)\citenamefont {Yamada},
  \citenamefont {Terao}, \citenamefont {Ishikawa}, \citenamefont {Yamaguchi},
  \citenamefont {Mitamura},\ and\ \citenamefont {Goto}}]{ya.te.91}%
  \BibitemOpen
  \bibfield  {author} {\bibinfo {author} {\bibfnamefont {H.}~\bibnamefont
  {Yamada}}, \bibinfo {author} {\bibfnamefont {K.}~\bibnamefont {Terao}},
  \bibinfo {author} {\bibfnamefont {F.}~\bibnamefont {Ishikawa}}, \bibinfo
  {author} {\bibfnamefont {M.}~\bibnamefont {Yamaguchi}}, \bibinfo {author}
  {\bibfnamefont {M.}~\bibnamefont {Mitamura}}, \ and\ \bibinfo {author}
  {\bibfnamefont {T.}~\bibnamefont {Goto}},\ }\href@noop {} {\bibfield
  {journal} {\bibinfo  {journal} {J. Phys: Condens. Matter}\ }\textbf {\bibinfo
  {volume} {11}},\ \bibinfo {pages} {483} (\bibinfo {year} {1991})}\BibitemShut
  {NoStop}%
\bibitem [{\citenamefont {Steinbeck}\ \emph {et~al.}(2001)\citenamefont
  {Steinbeck}, \citenamefont {Richter},\ and\ \citenamefont
  {Eschrig}}]{st.ri.01}%
  \BibitemOpen
  \bibfield  {author} {\bibinfo {author} {\bibfnamefont {L.}~\bibnamefont
  {Steinbeck}}, \bibinfo {author} {\bibfnamefont {M.}~\bibnamefont {Richter}},
  \ and\ \bibinfo {author} {\bibfnamefont {H.}~\bibnamefont {Eschrig}},\ }\href
  {\doibase 10.1103/PhysRevB.63.184431} {\bibfield  {journal} {\bibinfo
  {journal} {Phys. Rev. B}\ }\textbf {\bibinfo {volume} {63}},\ \bibinfo
  {pages} {184431} (\bibinfo {year} {2001})}\BibitemShut {NoStop}%
\bibitem [{\citenamefont {Campbell}(1972)}]{camp.72}%
  \BibitemOpen
  \bibfield  {author} {\bibinfo {author} {\bibfnamefont {I.~A.}\ \bibnamefont
  {Campbell}},\ }\href@noop {} {\bibfield  {journal} {\bibinfo  {journal} {J.
  Phys. F: Metal Phys. 2}\ }\textbf {\bibinfo {volume} {2}},\ \bibinfo {pages}
  {L47} (\bibinfo {year} {1972})}\BibitemShut {NoStop}%
\bibitem [{\citenamefont {Ochi}\ \emph
  {et~al.}(2015{\natexlab{b}})\citenamefont {Ochi}, \citenamefont {Arita},
  \citenamefont {Matsumoto}, \citenamefont {Kino},\ and\ \citenamefont
  {Miyake}}]{oc.ry.15}%
  \BibitemOpen
  \bibfield  {author} {\bibinfo {author} {\bibfnamefont {M.}~\bibnamefont
  {Ochi}}, \bibinfo {author} {\bibfnamefont {R.}~\bibnamefont {Arita}},
  \bibinfo {author} {\bibfnamefont {M.}~\bibnamefont {Matsumoto}}, \bibinfo
  {author} {\bibfnamefont {H.}~\bibnamefont {Kino}}, \ and\ \bibinfo {author}
  {\bibfnamefont {T.}~\bibnamefont {Miyake}},\ }\href {\doibase
  10.1103/PhysRevB.91.165137} {\bibfield  {journal} {\bibinfo  {journal} {Phys.
  Rev. B}\ }\textbf {\bibinfo {volume} {91}},\ \bibinfo {pages} {165137}
  (\bibinfo {year} {2015}{\natexlab{b}})}\BibitemShut {NoStop}%
\bibitem [{\citenamefont {Wills}\ \emph {et~al.}(2010)\citenamefont {Wills},
  \citenamefont {Alouani}, \citenamefont {Andersson}, \citenamefont {Delin},
  \citenamefont {Eriksson},\ and\ \citenamefont {Grechnev}}]{rsptbook}%
  \BibitemOpen
  \bibfield  {author} {\bibinfo {author} {\bibfnamefont {J.~M.}\ \bibnamefont
  {Wills}}, \bibinfo {author} {\bibfnamefont {M.}~\bibnamefont {Alouani}},
  \bibinfo {author} {\bibfnamefont {P.}~\bibnamefont {Andersson}}, \bibinfo
  {author} {\bibfnamefont {A.}~\bibnamefont {Delin}}, \bibinfo {author}
  {\bibfnamefont {O.}~\bibnamefont {Eriksson}}, \ and\ \bibinfo {author}
  {\bibfnamefont {O.}~\bibnamefont {Grechnev}},\ }\href@noop {} {\emph
  {\bibinfo {title} {Full-Potential Electronic Structure Method}}}\ (\bibinfo
  {publisher} {Springer, Berlin},\ \bibinfo {year} {2010})\BibitemShut
  {NoStop}%
\bibitem [{\citenamefont {Gr{\aa}n{\"a}s}\ \emph {et~al.}(2012)\citenamefont
  {Gr{\aa}n{\"a}s}, \citenamefont {Marco}, \citenamefont {Thunstr{\"o}m},
  \citenamefont {Nordstr{\"o}m}, \citenamefont {Eriksson}, \citenamefont
  {Bj{\"o}rkman},\ and\ \citenamefont {Wills}}]{gr.ma.12}%
  \BibitemOpen
  \bibfield  {author} {\bibinfo {author} {\bibfnamefont {O.}~\bibnamefont
  {Gr{\aa}n{\"a}s}}, \bibinfo {author} {\bibfnamefont {I.~D.}\ \bibnamefont
  {Marco}}, \bibinfo {author} {\bibfnamefont {P.}~\bibnamefont
  {Thunstr{\"o}m}}, \bibinfo {author} {\bibfnamefont {L.}~\bibnamefont
  {Nordstr{\"o}m}}, \bibinfo {author} {\bibfnamefont {O.}~\bibnamefont
  {Eriksson}}, \bibinfo {author} {\bibfnamefont {T.}~\bibnamefont
  {Bj{\"o}rkman}}, \ and\ \bibinfo {author} {\bibfnamefont {J.}~\bibnamefont
  {Wills}},\ }\href {\doibase 10.1016/j.commatsci.2011.11.032} {\bibfield
  {journal} {\bibinfo  {journal} {Comp. Mater. Sci.}\ }\textbf {\bibinfo
  {volume} {55}},\ \bibinfo {pages} {295–302} (\bibinfo {year}
  {2012})}\BibitemShut {NoStop}%
\bibitem [{\citenamefont {Perdew}\ and\ \citenamefont {Wang}(1992)}]{pe.wa.92}%
  \BibitemOpen
  \bibfield  {author} {\bibinfo {author} {\bibfnamefont {J.~P.}\ \bibnamefont
  {Perdew}}\ and\ \bibinfo {author} {\bibfnamefont {Y.}~\bibnamefont {Wang}},\
  }\href {\doibase 10.1103/PhysRevB.45.13244} {\bibfield  {journal} {\bibinfo
  {journal} {Phys. Rev. B}\ }\textbf {\bibinfo {volume} {45}},\ \bibinfo
  {pages} {13244} (\bibinfo {year} {1992})}\BibitemShut {NoStop}%
\bibitem [{\citenamefont {Schweizer}\ and\ \citenamefont
  {Tasset}(1980)}]{sc.ta.80}%
  \BibitemOpen
  \bibfield  {author} {\bibinfo {author} {\bibfnamefont {J.}~\bibnamefont
  {Schweizer}}\ and\ \bibinfo {author} {\bibfnamefont {F.}~\bibnamefont
  {Tasset}},\ }\href {http://stacks.iop.org/0305-4608/10/i=12/a=020} {\bibfield
   {journal} {\bibinfo  {journal} {J. Phys. F}\ }\textbf {\bibinfo {volume}
  {10}},\ \bibinfo {pages} {2799} (\bibinfo {year} {1980})}\BibitemShut
  {NoStop}%
\bibitem [{\citenamefont {Alameda}\ \emph {et~al.}(1981)\citenamefont
  {Alameda}, \citenamefont {Givord}, \citenamefont {Lemaire},\ and\
  \citenamefont {Lu}}]{al.gi.81}%
  \BibitemOpen
  \bibfield  {author} {\bibinfo {author} {\bibfnamefont {J.~M.}\ \bibnamefont
  {Alameda}}, \bibinfo {author} {\bibfnamefont {D.}~\bibnamefont {Givord}},
  \bibinfo {author} {\bibfnamefont {R.}~\bibnamefont {Lemaire}}, \ and\
  \bibinfo {author} {\bibfnamefont {Q.}~\bibnamefont {Lu}},\ }\href@noop {}
  {\bibfield  {journal} {\bibinfo  {journal} {J. Appl. Phys.}\ }\textbf
  {\bibinfo {volume} {52}},\ \bibinfo {pages} {2079} (\bibinfo {year}
  {1981})}\BibitemShut {NoStop}%
\bibitem [{\citenamefont {Nordstr\"om}\ \emph {et~al.}(1990)\citenamefont
  {Nordstr\"om}, \citenamefont {Eriksson}, \citenamefont {Brooks},\ and\
  \citenamefont {Johansson}}]{no.er.90}%
  \BibitemOpen
  \bibfield  {author} {\bibinfo {author} {\bibfnamefont {L.}~\bibnamefont
  {Nordstr\"om}}, \bibinfo {author} {\bibfnamefont {O.}~\bibnamefont
  {Eriksson}}, \bibinfo {author} {\bibfnamefont {M.~S.~S.}\ \bibnamefont
  {Brooks}}, \ and\ \bibinfo {author} {\bibfnamefont {B.}~\bibnamefont
  {Johansson}},\ }\href {\doibase 10.1103/PhysRevB.41.9111} {\bibfield
  {journal} {\bibinfo  {journal} {Phys. Rev. B}\ }\textbf {\bibinfo {volume}
  {41}},\ \bibinfo {pages} {9111} (\bibinfo {year} {1990})}\BibitemShut
  {NoStop}%
\bibitem [{\citenamefont {L.~X.~Benedict}(2015)}]{be.ab.15}%
  \BibitemOpen
  \bibfield  {author} {\bibinfo {author} {\bibfnamefont {P.~S. B. S. M.~D.}\
  \bibnamefont {L.~X.~Benedict}, \bibfnamefont {D.~Aberg}},\ }\href@noop {}
  {\bibfield  {journal} {\bibinfo  {journal} {LLNL-TR-678582}\ } (\bibinfo
  {year} {2015})}\BibitemShut {NoStop}%
\bibitem [{\citenamefont {Lichtenstein}\ and\ \citenamefont
  {Katsnelson}(1998)}]{li.ka.97}%
  \BibitemOpen
  \bibfield  {author} {\bibinfo {author} {\bibfnamefont {A.~I.}\ \bibnamefont
  {Lichtenstein}}\ and\ \bibinfo {author} {\bibfnamefont {M.~I.}\ \bibnamefont
  {Katsnelson}},\ }\href@noop {} {\bibfield  {journal} {\bibinfo  {journal}
  {Phys. Rev. B}\ }\textbf {\bibinfo {volume} {57}},\ \bibinfo {pages} {6884}
  (\bibinfo {year} {1998})}\BibitemShut {NoStop}%
\bibitem [{\citenamefont {Larson}\ \emph {et~al.}(2003)\citenamefont {Larson},
  \citenamefont {Mazin},\ and\ \citenamefont {Papaconstantopoulos}}]{la.ma.03}%
  \BibitemOpen
  \bibfield  {author} {\bibinfo {author} {\bibfnamefont {P.}~\bibnamefont
  {Larson}}, \bibinfo {author} {\bibfnamefont {I.~I.}\ \bibnamefont {Mazin}}, \
  and\ \bibinfo {author} {\bibfnamefont {D.~A.}\ \bibnamefont
  {Papaconstantopoulos}},\ }\href {\doibase 10.1103/PhysRevB.67.214405}
  {\bibfield  {journal} {\bibinfo  {journal} {Phys. Rev. B}\ }\textbf {\bibinfo
  {volume} {67}},\ \bibinfo {pages} {214405} (\bibinfo {year}
  {2003})}\BibitemShut {NoStop}%
\bibitem [{\citenamefont {Heidemann}\ \emph {et~al.}(1975)\citenamefont
  {Heidemann}, \citenamefont {Richter},\ and\ \citenamefont
  {Buschow}}]{he.ri.75}%
  \BibitemOpen
  \bibfield  {author} {\bibinfo {author} {\bibfnamefont {A.}~\bibnamefont
  {Heidemann}}, \bibinfo {author} {\bibfnamefont {D.}~\bibnamefont {Richter}},
  \ and\ \bibinfo {author} {\bibfnamefont {K.~H.~J.}\ \bibnamefont {Buschow}},\
  }\href@noop {} {\bibfield  {journal} {\bibinfo  {journal} {Zeitschrisft F\"ur
  Phys. B Condens. Matter}\ }\textbf {\bibinfo {volume} {22}},\ \bibinfo
  {pages} {367} (\bibinfo {year} {1975})}\BibitemShut {NoStop}%
\bibitem [{\citenamefont {Burzo}\ \emph {et~al.}(2011)\citenamefont {Burzo},
  \citenamefont {Chioncel}, \citenamefont {Tetean},\ and\ \citenamefont
  {Isnard}}]{bu.ch.11}%
  \BibitemOpen
  \bibfield  {author} {\bibinfo {author} {\bibfnamefont {E.}~\bibnamefont
  {Burzo}}, \bibinfo {author} {\bibfnamefont {L.}~\bibnamefont {Chioncel}},
  \bibinfo {author} {\bibfnamefont {R.}~\bibnamefont {Tetean}}, \ and\ \bibinfo
  {author} {\bibfnamefont {O.}~\bibnamefont {Isnard}},\ }\href@noop {}
  {\bibfield  {journal} {\bibinfo  {journal} {J. Phys: Condens. Matter}\
  }\textbf {\bibinfo {volume} {23}},\ \bibinfo {pages} {026001} (\bibinfo
  {year} {2011})}\BibitemShut {NoStop}%
\bibitem [{\citenamefont {Hong}\ \emph {et~al.}(1995)\citenamefont {Hong},
  \citenamefont {Hauser}, \citenamefont {Hilscher}, \citenamefont {Gratz},
  \citenamefont {Ballou},\ and\ \citenamefont {Ido}}]{ho.ha.95}%
  \BibitemOpen
  \bibfield  {author} {\bibinfo {author} {\bibfnamefont {N.}~\bibnamefont
  {Hong}}, \bibinfo {author} {\bibfnamefont {R.}~\bibnamefont {Hauser}},
  \bibinfo {author} {\bibfnamefont {G.}~\bibnamefont {Hilscher}}, \bibinfo
  {author} {\bibfnamefont {E.}~\bibnamefont {Gratz}}, \bibinfo {author}
  {\bibfnamefont {R.}~\bibnamefont {Ballou}}, \ and\ \bibinfo {author}
  {\bibfnamefont {H.}~\bibnamefont {Ido}},\ }\href {\doibase
  10.1016/0304-8853(94)00759-4} {\bibfield  {journal} {\bibinfo  {journal}
  {Journal of Magnetism and Magnetic Materials}\ }\textbf {\bibinfo {volume}
  {140-144}},\ \bibinfo {pages} {933–934} (\bibinfo {year}
  {1995})}\BibitemShut {NoStop}%
\end{thebibliography}%
\end{document}